 \documentclass{aa}  

\usepackage{graphicx}
\usepackage{txfonts}
\newcommand{\xmm}{\mbox{\em XMM-Newton}}
\newcommand{\chandra}{\mbox{\em Chandra}}
\newcommand{\hd}{HD~189733}

\begin{document}

   \title{No X-rays from WASP-18. 
}

   \subtitle{Implications for its age, activity, and the influence of its massive hot Jupiter}

   \author{
	Ignazio Pillitteri\inst{1,2}
	\and Scott J. Wolk\inst{2}
	\and Salvatore Sciortino\inst{1} 
	\and Victoria Antoci\inst{3} 
          }

\institute{
	Osservatorio Astronomico di Palermo, 
	Piazza del Parlamento 1, 90134 Palermo, Italy\\
	\email{pilli@astropa.inaf.it}
	\and
	SAO-Harvard Center for Astrophysics, 
	60 Garden St, Cambridge, MA, 02138, USA\\
	\and
	Stellar Astrophysics Centre, 
	Department of Physics and Astronomy, Aarhus University, 
	Ny Munkegade 120, DK-8000 Aarhus C, Denmark.
	}

\date{Received; accepted}

 
  \abstract{
{ About 20\% out of the $>1000$ known exoplanets are Jupiter analogs 
orbiting very close to their parent stars.}
It is still under debate to what detectable level such {\em hot Jupiters} 
possibly affect the activity of the host stars through tidal or magnetic 
star-planet interaction.
In this paper we report on an 87 ks {\em Chandra} observation of the hot Jupiter 
hosting star WASP-18. This system is composed of an F6 type star 
and a hot Jupiter of mass $10.4 M_{Jup}$ orbiting in less than 20 hr around the parent star.
{ On the basis of an isochrone fitting, WASP-18 is thought to be 600 Myr old 
{ and within the range of uncertainty of 0.5-2 Gyr.} } 
The star is not detected in X-rays down to a luminosity limit
of $4\times10^{26}$ erg/s, more than two orders of magnitude lower than expected for a star of 
this age and mass. This value proves an unusual lack of activity for a star with estimated age
around 600 Myr. We argue that the massive planet can play a crucial role in disrupting 
the stellar magnetic 
dynamo created within its thin convective layers.

Another additional 212 X-ray sources are detected in the \chandra\ image. 
We list them and briefly discuss their nature.
}

   \keywords{stars: activity -- stars: coronae -- stars: individual (WASP-18)}

   \maketitle
%

\section{Introduction}
{ About 20\%  of the $>1000$ extra-solar planets} discovered to date are hot Jupiters, 
meaning massive planets orbiting at a few stellar radii from the parent stars. 
Models predict that hot Jupiters could affect the activity of their host stars 
through either tidal or  magneto-hydrodynamical interaction (e.g. \citealp{Cuntz2000} and 
\citealp{Ip2004}). { Both effects strongly scale with the separation $d$
between the two bodies 
\citep{Saar2004}). }
Observational evidence for star-planet interaction (SPI) was first reported by \cite{Shkolnik03}, 
who discovered variability in the chromospheric activity indicators, the H\&K lines of Ca~II of HD~179949, 
phased with the planetary motion. Subsequently, \cite{Fares2010} reconstructed the 
magnetic field of HD~179949, and confirmed periodic variations of chromospheric activity indicators
(H$\alpha$ and Ca~II lines) { synchronized} with the beat period of the planet-star system.

In the X-ray band, \cite{Kashyap08} showed that stars with hot 
Jupiters are statistically brighter by up to a factor four than stars with distant planets.
\cite{Krejcova2012} support the findings of Kashyap et al. (2008)
by means of a survey of Ca~II H\&K lines on planet hosting stars that traces a relationship
betweeen stellar activity and planet-star separation, with closer hot Jupiters orbiting
more active stars. 
However, \cite{Poppenhaeger2010}  and \cite{Poppenhaeger2011}
argue that that the above results are biased by selection effects, and  
SPI is not a common phenomenon, rather SPI manifests itself only in peculiar cases. 

One of the clearest cases of detection of SPI is \hd.
We have observed this system three times with \xmm\ and found strong evidence of SPI at 
work in X-rays (\citealp{Pillitteri2010, Pillitteri2011}, Pillitteri et al. 2014, submitted to ApJ). 
The first evidence comes from the overall activity of the host star.
The very low level of X-ray activity of the M type companion, \hd B, puts a strong constraint 
on the age of the system at $\ge 2$ Gyr \citep{Pillitteri2010,Poppenhaeger2013}. 
Supported also by the 2011 and 2012 \xmm\ observations, the old age of \hd B is reinforced
by the fact that it does not show flaring variability { on a time scale of a few
hours and within the \xmm\ exposures ($30-50$ ks)}, as in young or active M-type stars. 
The old age of the secondary is inconsistent with the young age of the system 
as derived from stellar activity of \hd A, which is of order 600 Myr \citep{Melo2006}.
\cite{Schroeter2011} have reported a similar result on Corot-2A.
Based on \chandra\ observations of a planetary transit, they find that the primary is X-ray bright 
with a luminosity $\sim1.9\cdot 10^{29}$ erg s$^{-1}$, 
indicating an age $<$ 300 Myr,  while a potential stellar companion of Corot~2A is undetected 
down to a limit of  $L_X \sim9\cdot10^{26}$ ergs s$^{-1}$, implying a much older age.

In main sequence stars, coronal activity is tightly linked with the internal structure of the stars, 
because of the link between convective zone, dynamo action and magnetic field emergence. 
Stellar activity is a handle for understanding the depth of the convective zone 
and the efficiency of the dynamo. In stars with intermediate masses, approximately from late A-type stars
and moving toward earlier types, the thin convective layer disappears, 
and so do magnetic dynamo and coronal emission. 
The precise onset of the convection is a function of the mass, the chemical composition, which affects
the opacity of the inner layers of the star, and the stellar evolutionary stage. 
Mid F-type stars like our target, WASP-18, should possess a thin convective layer that can still
generate an $\alpha-\omega$ dynamo similar to the solar one 
and produce a X-ray bright corona. In fact, mid and late F stars
are X-ray bright in young clusters like the Hyades (600 Myr) at a level of $L_X >10^{28.5}$ erg/s. 
In the framework of the relationship between age, rotation, convective zone depth, 
dynamo and coronal activity, the emission of X-rays in this range of masses is a probe of
the dynamo efficiency from  thin convective layer coupled with
the rotation of the star.  In late F-type stars with hot Jupiters at the age of Hyades,
an enhancement of the X-ray activity should be then expected. 

\subsection{WASP-18}
With the intent of exploring SPI as a function of the star-planet separation in a star with shallow convective 
zone, we observed the system of WASP-18 with {\em Chandra} for a whole orbital period of its
planet. WASP-18 (HD~10069, 2MASS J01372503-4540404) is a F6 star at $\sim100$ pc from the Sun 
that harbors a very close-in hot Jupiter, with 
the planet orbiting in only 19.4 hours \citep{Hellier2009}.
The main characteristics of this system are given in Table \ref{wasp18:table} and are obtained
from a recent spectroscopic study by \citet{Doyle2013}. 
Numerous  optical spectroscopic observations allowed quite precise estimates  
of the effective temperature, gravity, distance, chemical abundances and mass of WASP-18 
\citep{Doyle2013}. 

WASP-18b has a mass of about $10.43 M_{jup}$ and a density 
$\rho = 6.6 \rho_{Jup}$. 
Due to a star-planet separation of only 3.48 $R_*$ (0.02047 AU), 
the planet is experiencing strong irradiation that heats and bloats 
its atmosphere and  fills its Roche lobe. \cite{Southworth2009} estimate an 
equilibrium temperature of about 2400 K. The close separation suggests that the planet is on 
the verge of the final spiralling phase toward the parent star and this gives an  
opportunity to observe the final phases of a planet before destruption \citep{Brown2011}.

{ While WASP-18 was known to have low activity based on the $\log R^\prime_{HK}$ indicator, 
the remarkably close separation between planet and star demands further studies of this system
to explore the effects of SPI, both of tidal and magnetical origin. 
A star with an age of 600 Myr is expected to have a level
of X-ray luminosity typical of that of Hyades ($L_X \sim 10^{28.5}-10^{29.5}$ erg/s, 
\citealp{Stern1995,Randich1995}). 
Despite this range of presumed X-ray luminosity, WASP-18 was undetected in a 50ks stacked {\em Swift} 
exposure \citep{Miller2012} at a limit of $\log L_{X,lim} = 27.5$, and in the 
ROSAT All Sky Survey (which has an typical sensitivity of $\log f_X \ge10^{-12}$ 
erg s$^{-1}$ cm$^{-2}$).
}
  
In this paper, we report the absence of detected X-ray emission from WASP-18 in a 87 ks deep
\chandra\ exposure, and its implications
for the models of star-planet interaction and the evolutionary stage of this system. 
The structure of the paper is as follows: Sect. 2 describes the information on the age of WASP-18,
Sect. 3 describes the observations and data analysis. Sect. 4 reports our results.
Finally, In Sect. 5 and 6 we discuss the results and draw our conclusions, respectively.

\section{Age of WASP-18}
{ 
With respect to our investigations the age of WASP-18 is a critical parameter, 
hence in this section. we discuss it in details.
Estimating the age of a star is a difficult task, and best applied to a 
statistical sample like stars belonging to open clusters. 
Methods for dating the age of stars are semi- or fully empirical and rely 
on gyrochronology, rotation and activity tracers, or on isochrone fitting 
or asteroseismology, this latter especially used for solar-like oscillators and red giants  
\citep{Soderblom2010}. 
It is observed that stars during the main sequence phase lose angular momentum through
magnetized stellar winds so that for single stars it is possible to estimate their 
age from their rotation \citep{Skumanich1972}. 
Historically, activity tracers have been the Ca II H\&K lines, Mg II h\&k lines and H$\alpha$ line
\citep{Wilson1966,Vaughan1980,Baliunas1995}. 
All of them are sensitive to the chromospheric contribution to the line that is related to the overall
stellar activity. Again, the connection between activity and rotation and between 
rotation and age makes the measurements of these lines an empirical method to estimate the 
stellar age. However, in the cases of stars with hot Jupiters, the rotation and the activity 
tracers can be affected by the interaction with the planet, thus biasing the estimate of the age
of the host star. 
\citet{Pillitteri2010,Pillitteri2011,Poppenhaeger2013,Pillitteri2014,Poppenhaeger2014} and 
\citet{Schroeter2011} have found that the hot Jupiter hosting stars HD~189733 and Corot-2A
have likely been spun up by their close in planets, and thus their activity and rotation have
been enhanced, mimicking thus the behavior of younger stars. 
In HD~189733, activity tracers like X-ray luminosity would assign
an age in the range 0.6-1.1 Gyr \citep{Melo2006,Sanz-Forcada2011}, 
while the stellar companion is much older. { On the same star,  
\citet{Torres2008} used fitting to $Y^2$ isochrones \citep{Demarque2004,Yi2001} 
to estimate an age of $\tau =6.8_{-4.4}^{+5.2}$ Gyr.}
 
{ The low chromospheric activity of WASP-18 would assign to it an age 
similar to that of the Sun or older.} 
On the basis of isochrone fitting, the age of WASP-18 has been 
estimated by \citet{Hellier2009} and \citet{Southworth2009} to be similar to that of the 
Hyades but this estimate has a large range of uncertainty. 
\citet{Southworth2009} studied in details the stellar parameters of WASP-18 employing several
models of stellar evolution: models from Claret and collaborators 
\citep{Claret2004,Claret2005,Claret2006,Claret2007}, $Y^2$ models
\citep{Demarque2004,Yi2001}, and {\em Cambridge} models 
\citep{Eldridge2004, Pols1998}. The parameters from various models agree well 
(less agreement is found for the results from {\em Cambridge} models) but overall the age of
WASP-18 is found in the range from 0.5 to at most 2 Gyr.
We will assume this range of age for WASP-18 and in Sect. \ref{results} 
we will compare these values against the evidences of an older stellar age.
For a star in the 0.5-2 Gyr age time interval, corresponding to an age between that of the Hyades 
and of stars in NGC~752 open cluster, the X-ray luminosity of late F stars should be approximately in the range
$28.1< \log L_X < 29.5$ \citep{Pallavicini81,Stern1995,Randich1995,Giardino07}.

}
\begin{table*}
\caption{\label{wasp18:table} Main properties of the WASP-18 system. 
Photospheric data from \citet{Doyle2013},
other data from the catalog at {\em exoplanet.eu}. 
The range of age is  from \citet{Southworth2009}}
\resizebox{\textwidth}{!}{
\begin{tabular}{c c c c c c c c}\hline \hline
Name & Type & Mass & Radius & Distance & $T_\mathrm{eff}$ & V & Age \\
WASP-18 & F6IV-V & $1.28\pm0.09$ M$_\odot$ & $1.29\pm0.16$ R$_\odot$ & $100\pm10$ pc & $6400\pm75$ K & 9.3 mag & $0.63$  ($0.5-2$ Gyr) \\
\hline 
Name & Type & Mass & Radius & Period & Semi-major Axis & Note \\ 
WASP-18b & hot Jupiter & $10.4\pm0.4$ M$_{Jup}$ & $1.165\pm0.077$ R$_{Jup}$ & $0.9414518\pm4\times10^{-7}$ d & $0.0205\pm0.0004$ AU & Transiting \\
\hline
\end{tabular}
}
\end{table*}

\section{Observation and data analysis}
\begin{figure}
 \begin{center}
 \includegraphics[width=\columnwidth]{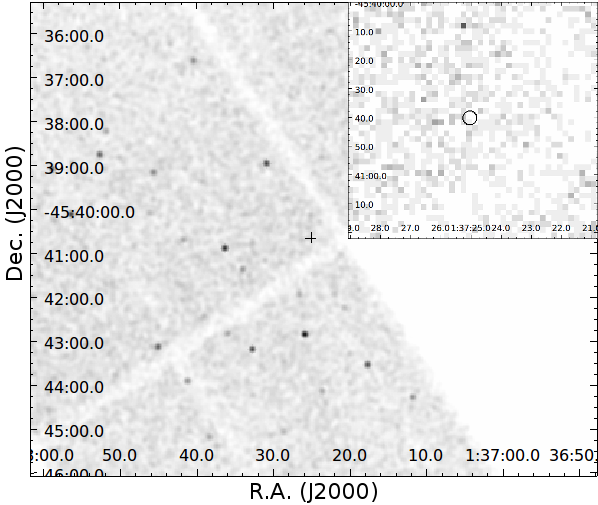}
\end{center}
\caption{\label{acis-img}
{\em Chandra}-ACIS image toward WASP-18. The cross marks the position
reported in SIMBAD database. The inset image at the top right corner shows
the region around the star, with the circle marking the position of WASP-18.}
\end{figure}

We observed WASP-18 ($\alpha=1^h37^m24.2^s$, $\delta=-45^d40^m40.3^s$, J2000) 
using \chandra\ with a continuous $\sim87$ks long observation with ACIS camera.
ACIS CCDs number 1, 2, 3, 6, 7 were used. WASP-18 falls in the CCD nr. 3 (see Fig. \ref{acis-img}). 
The star  is not visible in the X-ray image, and is not detected at a significance threshold of $4 \sigma$ of local background, 
after applying a wavelet detection algorithm \citep{Damiani97a, Damiani97b,Damiani2003}.
The significance threshold of $4 \sigma$ is the value usually adopted to have statistically at most one 
spurious source per field. 
We have also run the detection algorithm at a significance threshold of 3$\sigma$ but still no sources
are found within 5\arcsec\ from the nominal position of the star. 
{ We tested the hypothesis that WASP-18 could be heavily embedded in the material stripped from the outer
atmosphere of the planetary companion. In this case the gas absorption should attenuate the soft part
of the spectrum ($kT< 1$ keV), leaving any hard component spectrum still detectable. 
For this purpose we applied 
the same source detection procedure on the image in the 1.5-5.0 keV band without successful detection 
of any source at the position of WASP-18. Hence we exclude that WASP-18 is shrouded in X-rays 
by a dense gas layer from its hot Jupiter.} 
At this point, we have calculated an upper limit to the rate in 0.3-8.0 keV consistent with the 
3$\sigma$ threshold, this value is $3.8\times10^{-5}$ counts s$^{-1}$. 

Table \ref{listasrc} reports the positions, offaxis, significance, counts,
count rates, and exposure times of 212 detected sources detected above the $4 \sigma$ threshold. 
We have cross-correlated the positions of the detected sources against NED, 2MASS and Simbad catalogs
in order to identify optical and IR counterparts. Table \ref{opt-id} reports the list of matches with
notes about their nature.
For the brightest sources we have also extracted the spectra, and performed a best fit adopting 
either an absorbed thermal model or an absorbed power law with XSPEC software ver. 12.8.
Table \ref{fit} lists the best fit parameters. 

{ We obtained an archival FEROS spectrum  of WASP-18 acquired on 
Sept. 9th 2010, in particular we examinated the portion of 
spectrum around $H_\alpha$ and Lithium doublet (6708\AA).
We estimated the rotational period of WASP-18 by deriving the projected rotational velocity v~sin~$i$.
We measured the gaussian full width at half maximum of the weak Fe lines around Li doublet equal to 
$FWHM = 0.42$ \AA.
We corrected this value by quadratically subtracting: a) 0.14\AA\ of instrumental resolution, 
b) 1 km/s of micro turbulence, and c) 4 km/s of macro turbulence. { The values 
of micro and macro turbulence are broadly consistent with those used by \citet{Doyle2013}.}
We assumed that the axis of the planetary orbit aligned with the stellar rotational
axis and a angle between orbital plane and line of sight $i = 86\pm2.5 \deg$  to obtain a value
of rotational velocity $v \sim 17.2\pm0.5$ km/s. Given a stellar radius of 1.29 R$_\odot$, 
the rotational period is $3.7-3.9$ days, which makes WASP-18 a moderate rotator.
However, \citet{Doyle2013} report  a slower v~sin~$i$ ($10.9\pm0.7$ km/s) which would give a 
period of $P\sim6$ days. { The discrepancy of our period (3.7-3.9 d) and that inferred by Doyle et  al. (2013) does not produce disagreement in the expected activity of WASP-18.}
For a star of the mass of WASP-18 and rotating in about $4-6$ days, 
the expected X-ray luminosity  should be $L_X\ge 10^{29}$ erg s$^{-1}$ 
\citep[see Fig. 5 and Fig. 8 in][]{Pizzolato2003}. 
}

\begin{table*}
\caption{ List of X-ray sources detected in the ACIS image. Only ten rows are shown here, 
the full table is available in electronic format only. \label{listasrc}}
\begin{center}
\resizebox{0.95\textwidth}{!}{
\begin{tabular}{r r r r r r r r r r r}
\hline\hline
  \multicolumn{1}{c}{num.} &
  \multicolumn{1}{c}{R.A. (J2000)} &
  \multicolumn{1}{c}{Dec} (J2000)&
  \multicolumn{1}{c}{Pos. err} &
  \multicolumn{1}{c}{Off-axis} &
  \multicolumn{1}{c}{Significance} &
  \multicolumn{1}{c}{Counts} &
  \multicolumn{1}{c}{Cts err.} &
  \multicolumn{1}{c}{Rate} &
  \multicolumn{1}{c}{Rate err.} &
  \multicolumn{1}{c}{Exp. time} \\
& deg & deg & arcsec & arcmin & $\sigma_{bkg}$ & \multicolumn{2}{c}{cts} &  \multicolumn{2}{c}{ct ks$^{-1}$} & ks \\  
\hline
1	 & 	24.23177	 & 	-45.54562	 & 	8.4	 & 	9.78	 & 	4.45	 & 	69.51	 & 	21.66	 & 	1.99	 & 	0.62	 & 	34.93  \\
2	 & 	24.24039	 & 	-45.58173	 & 	2.5	 & 	7.77	 & 	4.13	 & 	18.18	 & 	7.29	 & 	0.296	 & 	0.119	 & 	61.37  \\
3	 & 	24.24631	 & 	-45.52741	 & 	3.3	 & 	10.46	 & 	4.38	 & 	24.7	 & 	9.64	 & 	0.79	 & 	0.308	 & 	31.25  \\
4	 & 	24.25307	 & 	-45.62677	 & 	1	 & 	5.41	 & 	9.44	 & 	30.27	 & 	9.06	 & 	0.399	 & 	0.119	 & 	75.95  \\
5	 & 	24.26107	 & 	-45.64003	 & 	0.8	 & 	4.66	 & 	4.77	 & 	7.54	 & 	4.5	 & 	0.123	 & 	0.073	 & 	61.51  \\
6	 & 	24.26128	 & 	-45.59062	 & 	1.9	 & 	6.83	 & 	4.2	 & 	13.33	 & 	5.74	 & 	0.183	 & 	0.079	 & 	72.66  \\
7	 & 	24.26154	 & 	-45.59641	 & 	2.5	 & 	6.54	 & 	4.39	 & 	18.43	 & 	7.22	 & 	0.249	 & 	0.097	 & 	74.09  \\
8	 & 	24.26997	 & 	-45.53962	 & 	2.2	 & 	9.4	 & 	12.45	 & 	76.95	 & 	11.65	 & 	1.359	 & 	0.206	 & 	56.61  \\
9	 & 	24.27015	 & 	-45.62587	 & 	1.4	 & 	4.93	 & 	8.56	 & 	32.49	 & 	9.56	 & 	0.409	 & 	0.12	 & 	79.52  \\
10	 & 	24.27129	 & 	-45.58701	 & 	2	 & 	6.79	 & 	7.52	 & 	35.92	 & 	10.77	 & 	0.487	 & 	0.146	 & 	73.81  \\
 \hline\end{tabular}
 }
 \end{center}
 \end{table*}
 
\section{\label{results} Results}
\subsection{No X-rays from WASP-18}
As stated in Sect. 2, WASP-18 is not detected in X-rays. The limiting count rate is 
$r_{lim} = 3.8\times10^{-5}$ ct s$^{-1}$.
By using PIMMS software we can estimate a limiting flux and luminosity from the upper limit to the rate.
We assumed a thermal spectrum with one temperature at $kT = 0.5$ keV in analogy with \hd\ and 
the plasma temperatures typical of young Hyades, and a gas absorption equal to  $N_H=1.5\times10^{20}$ cm$^{-2}$. 
We obtain a limit unabsorbed flux of $3.8\times10^{-16}$ erg s$^{-1}$ cm$^{-2}$
in 0.3-8.0 keV band, which corresponds to a luminosity of $L_X \le 4.5\times 10^{26}$ erg s$^{-1}$ at 
a distance of 100 pc. 
In the case we use a softer spectrum ($kT = 0.3 keV$) the limiting flux and luminosity would
be $f_X = 5.9\times10^{-16}$ erg s$^{-1}$ cm$^{-2}$ and $L_X = 1.5\times10^{27}$ erg s$^{-1}$,
respectively. 
The $L_X$ value are surprisingly low for a star whose age is estimated to be around 
600 Myr like the Hyades. 
late F stars in Hyades have a median luminosity of $L_X = 10^{29}$ erg s$^{-1}$ with a 
dispersion of about 0.5 dex \citep{Randich1998,Stern1995,Randich1995}.

Even older F type stars can emit more X-ray luminosity than WASP-18. 
For comparison, we report the serendipitous detection in X-rays of the F6V type star 
HD~110450, in a 20ks exposure devoted to observe R~Mus 
(P.I. N. Remage Evans, Pillitteri et al. in prep). 
HD~110450 is very similar to WASP-18: it is at $\sim$100 pc from the Sun, and
has an age estimated between 2.1 Gyr and 3.9 Gyr \citep{Casagrande2011,Holmberg2009}.
In X-rays, it has a PN rate of 20 ct/ks in 0.3-8.0 keV band, the best fit to the spectrum
gives a temperature of $kT = 0.2$ keV and $N_H = 3\times10^{21}$ cm$^{-2}$, unabsorbed
flux of $f_X = 2.5\times10^{-13}$ erg s$^{-1}$ cm$^{-2}$ 
and luminosity $L_X = 3.1 \times10^{29}$ erg s$^{-1}$. From this comparison, we conclude that
were WASP-18 coeval of HD~110450, its low X-ray activity would be still at odds with that of a typical 
F-type star. The inactive F5 type star Procyon and the F7 type hot Jupiter host 
$\tau$ Boo have an X-ray luminosities $\log L_X\sim 28$ and $\log L_X =28.8$ respectively.
 More interestingly, WASP-18 and $\tau$ Boo have similar rotational periods, 
{ 3.7$-$6} days for WASP-18, and 3.1-3.7 days for $\tau$ Boo \citep{Catala2007}. 
A similar level of X-ray activity should be observed 
because of the link between rotation and X-ray activity \citep{Pallavicini81,Pizzolato2003}.
{ However, the age of $\tau$ Boo is estimated around 2 Gyr, and its activity is perhaps b
oosted by magnetic SPI \citep{Walker2008}.}
WASP-18 stands out of the typical activity of similar mass stars.
Summarizing, WASP-18 is more than 2.5 orders of magnitude less luminous than Hyades, 
and about 2 orders of magnitude less active than the twin HD~110450, the inactive Procyon
and the similar planet host $\tau$ Boo.
The absence of X-ray activity in WASP-18 is in agreement with the very low 
chromospheric activity reported by \citet{Knutson2010} and  \citet{Miller2012}, 
that report the not detection of this star in 
a 50 ks stacked {\em Swift} exposure at $\log L_{X lim} = 27.5$.
Its low activity would suggest a much older age, of a few Gyr, because of the relationship between 
activity-rotation-age in main sequence stars,  and hence an age comparable to that of the Sun or older. 
{ Yet, the isochronal age estimate of WASP-18 appears plausible given the agreement of 
several stellar models \citep{Southworth2009}.  At odds with an age older than 2 Gyr is also the 
moderately high rotational rate of WASP-18.}
The contradicting age estimates of WASP-18 imply 
that even an old age alone cannot explain the darkness of WASP-18 in X-rays. 
This points out to a role played by its close massive  hot Jupiter.

%

\section{Discussion}
\begin{figure}
\begin{center}
\includegraphics[width=0.9\columnwidth,angle=0]{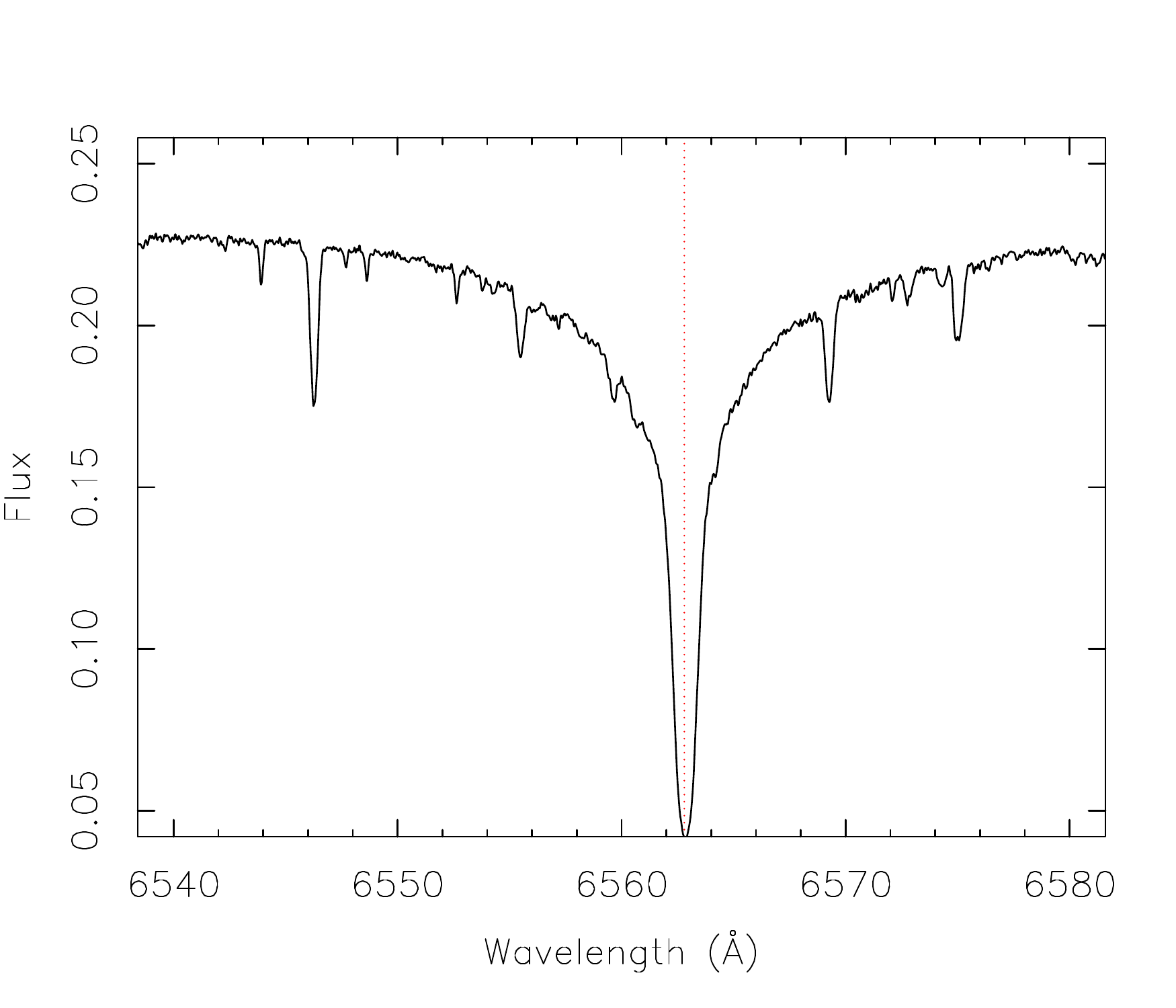}
\end{center}
\caption{\label{halpha-feros}
Portion of FEROS spectrum of WASP-18 around H$\alpha$ from the ESO 
archive of pipeline reduced spectra. 
This particular spectrum was taken on Sept 19th 2010. 
We use arbitrary units for flux on Y axis.
Absence of core emission in H$\alpha$, and narrow lines
support the idea that the star is not as young as Hyades and is a moderately high rotator.}
\end{figure}

The first unexpected result we have obtained with the 87 ks \chandra\ exposure is the very 
low upper limit to the X-ray luminosity of WASP-18 ($L_{X lim}=4.5\times10^{26}$ erg/s. 
The lack of X-ray and chromospheric activity of WASP-18 is inconsistent with its young age.
Indeed the comparison with Hyades and with the similar stars, like HD~110450, $\tau$ Boo 
and Procyon, points to an age much older than that given by \citet{Hellier2009}.  
Optical spectra obtained at ESO telescope with the FEROS spectrograph corfirm absence 
of activity and suggest { an age similar or older than that of the Sun} \citep{Soderblom2010}. 
Fig. \ref{halpha-feros} shows the portion of one of the FEROS spectra around H$\alpha$. 
The line is seen completely in absorption, with no signs of filling-in of the core that could be 
due to chromospheric activity. Moreover, the many weak absorption lines of Fe and other metals 
are quite narrow meaning a slow stellar rotation.
Altogether these features demonstrate the low activity of WASP-18, and would suggest an age 
more closer to that of the Sun. Using the empirical calibration given
by \citet{Soderblom2010}, despite its limitations, and the value of $\log R\prime _{HK} = -5.43$ reported
by \citet{Knutson2010}, gives us an age $\tau \ge 2.7$ Gyr. The chromospheric activity indicator
of WASP-18 is even below the average value of M67 cluster (4.5 Gyr, \citealp{Soderblom2010}) 
and the solar value. 

\begin{figure}
\begin{center}
\includegraphics[width=0.9\columnwidth,angle=0]{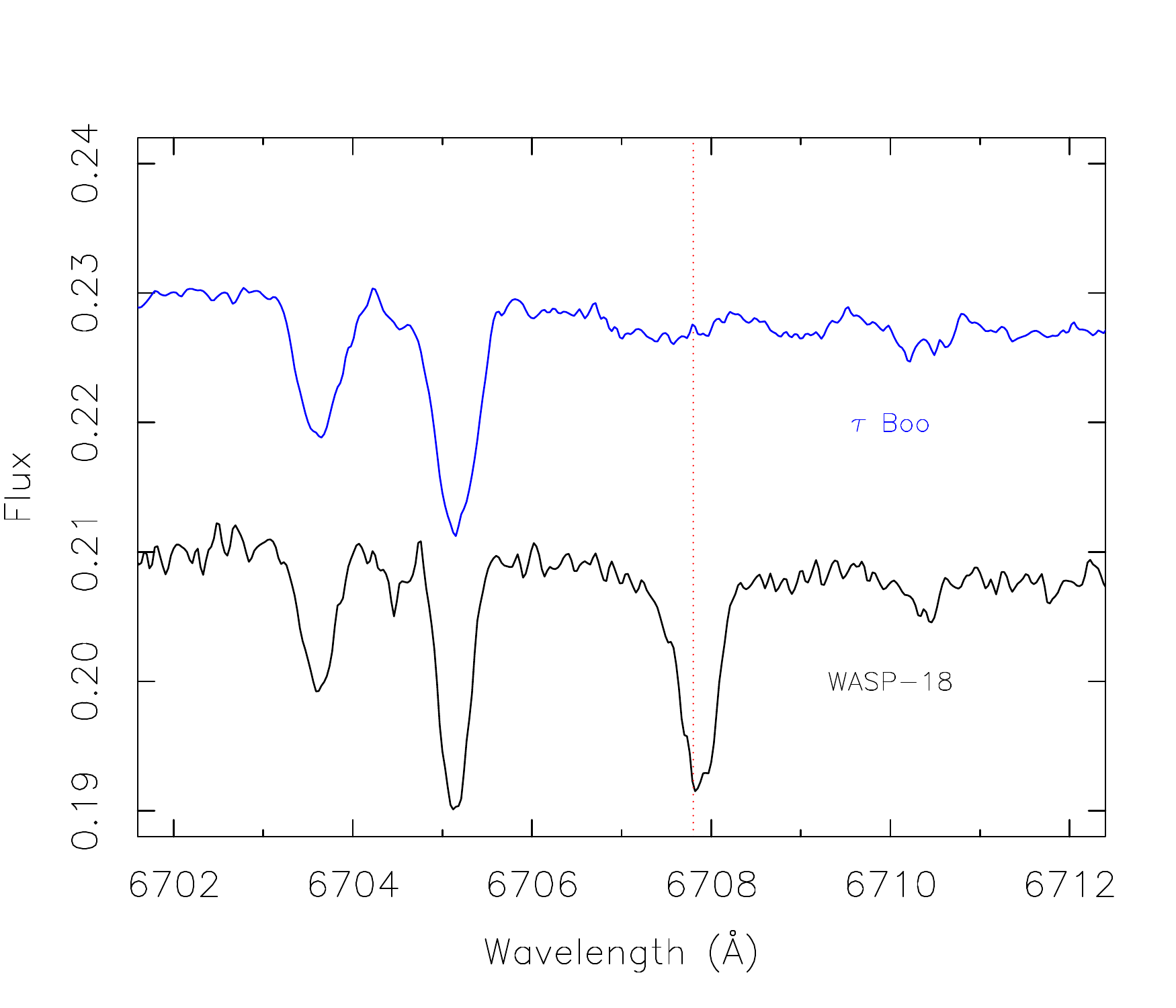}
\end{center}
\caption{\label{lithium-feros}
Portion of FEROS spectrum of WASP-18 around Li doublet at 6707.8\AA . 
This particular spectrum has been taken on Sept 19th 2010. 
We use arbitrary units for flux on Y axis.
For comparison we plot also the FEROS spectrum of $\tau$~Boo in the same 
spectral range. The spectrum of $\tau$~Boo has been scaled and shifted in 
wavelength of +0.42\AA\ (accounting for its radial velocity)
for an easier comparison with WASP-18.
The Li feature is strong in WASP-18 and absent in $\tau$~Boo.}
\end{figure}

\begin{figure}
\begin{center}
\includegraphics[width=0.9\columnwidth,angle=0]{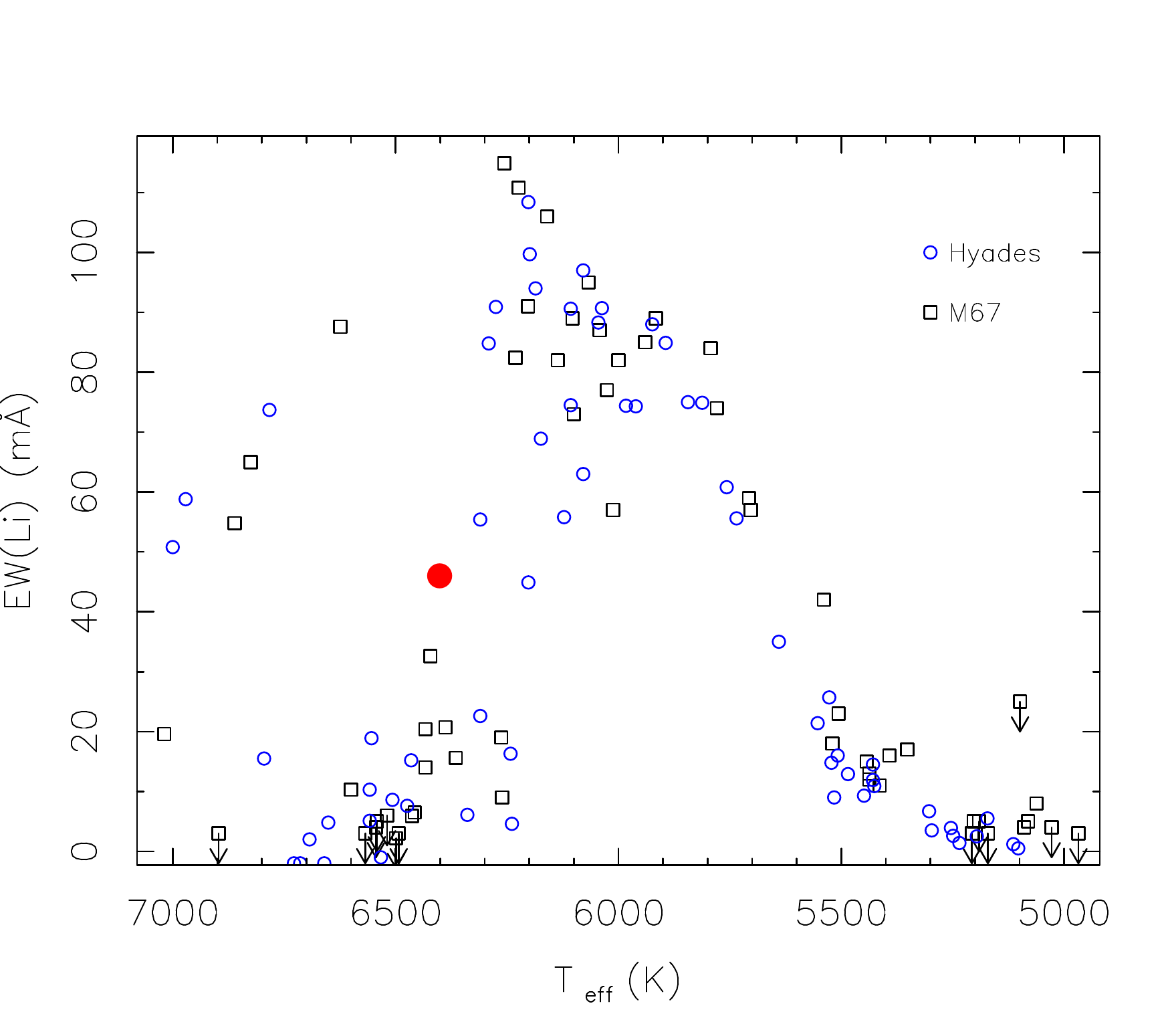}
\end{center}
\caption{\label{li_ew_teff}
EWs of Li vs. $T_{eff}$ for Hyades (open squares), M67 (open circles) and WASP-18 (filled circle).   
The EW(Li) of WASP-18 is $EW = 46\pm2$ m\AA\ and suggests a young age like Hyades, 
in contradiction with the absent activity of its corona and chromosphere.}
\end{figure}

However, the scenario becomes more puzzling when considering that WASP-18 shows
Li absorption at 6708\AA, with a value typical, or even stronger than values found in
F-late stars of Hyades \citep{Takeda2012} and M67 \citep{Pace2012}.
Fig. \ref{lithium-feros} shows a portion of the spectrum of WASP-18 around the Li doublet lines
at 6708 \AA, and Fig. \ref{li_ew_teff} shows the Li equivalent widths (EWs) vs. T$_{eff}$  given by 
\cite{Pace2012} and \cite{Takeda2012} for Hyades and the older M67 cluster. 
{ For comparison we show the FEROS spectrum of $\tau$~Boo which has no Li absorption.
The Li absorption in WASP-18 suggests a younger age with respect to  $\tau$~Boo or a different efficiency
in the mixing mechanism among these two stars of similar effective temperatures and masses.} 
In WASP-18 the equivalent width (EW) that we have measured is $EW(Li) = 46 \pm 2$ m\AA. 

Compared to the values reported by \cite{Takeda2012} and \cite{Pace2012}, 
WASP-18 has a slightly higher Li abundance. Lithium depletion is due to the convective mixing 
during the main sequence life of a star hence Li abundance is a rough indicator of youth in 
solar type stars with deep convective zone. 
In more massive stars, the mixing is less effective in bringing Li at the 
burning temperatures ($2.5\times10^6$ K). However in mid F type stars Li abundance shows a
significant dip that  is still not well understood.  
The dip appears between 6700 K and 6200 K (see Fig. \ref{lithium-feros}), 
with a steep edge on the hot side and a slower rise on the cool side.
WASP-18 is in this range of temperatures ($T_{eff}\sim 6400$ K) and should have a low
Li abundance. 

{ \citet{Israelian2004} claim that stars harboring hot Jupiters have Li abundance 
lower than single stars, similar results were reported by \citet{Gonzalez2008} 
and \citet{DelgadoMena2014}.
\citet{Bouvier2008} qualitatively explains this result by tracing it back to the Pre Main Sequence  (PMS)
history of the  angular momentum of the star+disk system. A long lived circumstellar disk
during PMS creates a slow stellar rotator with a strong shear at the base of the convective zone
and a more efficient mixing that accelerates the Li burning. At the same time, a long lived disk
offers more favorable conditions for the formation and the migration of exoplanets. 
WASP-18 has a significant Li abundance conflicting with the general pattern of  Li in stars hosting
hot Jupiters. } 
 
How to reconcile the X-ray darkness, the absence of activity, an old age and 
with the ``high" Li abundance in WASP-18? 
The solution to these conflicting evidences could rely in the strong tidal interaction between the 
massive planet and its host star. We speculate that the tidal interaction
in WASP-18 could interfer in a significant way with the upper layers of the convective zone 
to the point to cancel out the magnetic activity and to reduce the mixing of the stellar material.
Following \cite{Cuntz2000}, we estimated a height of the tide of $H_t \sim 498$ km.
This value could still be a small fraction of the depth of the convective zone 
($\sim16\%R_*$, \citealp{Houdek1999,Trampedach2013}). However, the depths of the convective zone 
in stars of mass of  WASP-18 or higher are very sensitive to the mass and temperature
and in this range of temperatures its calculation suffers of significant uncertainties 
(see \citealp{Claret2004}). 
WASP-18 has the large ratio between tide height and pressure scale height ($H_t/H_P$), 
being of order of 1.2 as shown in table \ref{hpress}, where we list the ratios $H_t/H_P$ of 
a sample of stars with hot Jupiters
with effective temperatures $T_{eff}$ in the range $6200-6600$ K taken from \citet{Knutson2010}. 
{ The  $H_t/H_P$ ratio takes into account the tidal effect due to the mass of the planet and 
its distance from the star (through $H_t$) and the properties of the star ($T_{eff}$, stellar 
mass and radius through $H_P$). 
We speculate that the tides on the stellar surface could influence the convective motions and the
meridional circulation inside the convective layers to effectively reduce or nullify 
the mechanism of magnetic dynamo.
The ratio $H_t/H_P$ could be an empirical parameter of the efficiency of the planetary tide in
reducing the shear within convective layers. The difference of Li in WASP-18 and $\tau$~Boo could be 
a manifestation of different tidal interactions in these two systems. 
It has been observed that in tidally locked binaries of Hyades Li is more abundant 
than in single stars pointing to a role of tidal influence on the inner mixing of these
stars \citep{Thorburn1993,Deliyannis1994}. 
}
%
%

The existence of WASP-18b poses a strong constraint on the models of the dynamics of planets 
migrating inward. 
If the stellar age is in the range 2-4 Gyr, the inward planet migration has acted on a time 
scale of a few Gyr, not on a time scale of hundreds of Myr as derived by \cite{Brown2011}. 
Hence the orbital evolution of WASP-18~b  has been slower than predicted by models
of orbital evolution of hot Jupiters.
%
The low activity of the star has also consequences for the photo-evaporation of the planet
and its lifetime. {An X-ray and UV flux two orders of magnitude weaker than in other systems like
$\tau$~Boo and \hd\  produce much less evaporation of the upper layers of
the planetary atmosphere, making the process slower than in other active hosts 
of hot Jupiters \citep{Penz08,Sanz-Forcada2011}. We expect that Roche-Lobe enhancement of the
evaporation \citep{Erkaev2007} should not be important in WASP-18b given its mass ($\sim10.4 M_{jup}$). }

The absence of a significant corona explains why the star is dark in X-rays. 
%
WASP-18 demonstrates that SPI of tidal and magnetic origin must depend on both stellar 
structure and evolutionary stage.
{ A different example of SPI in a star with a very close-in hot Jupiter is given by WASP-19. 
Like WASP-18, WASP-19 has a close-in hot Jupiter that orbits in less than one day. 
However, WASP-19 is a K type star with a deep convective zone and has a planet of almost the mass
of Jupiter. WASP-19 shows high chromospheric emission, at a level similar to \hd\ 
\citep{Knutson2010}. The ratio $H_t/H_P$ in WASP-19 is 18\% and the height of the tide 
is 55.2 km, a very small fraction of the depth of the convective zone in this star.
We argue that the different stellar structure of WASP-19 and WASP-18 likely results in different dynamo 
strength and coronal emission. We argue that in the case of WASP-19 the tidal interaction  
cannot affect significantly the motions of material inside the convective zone as in the case 
of WASP-18. As a consequence, a magnetic dynamo can be still established and magnetic SPI
is at work in this system enhancing the overall activity of WASP-19. With opposite effects, the tidal
interaction of WASP-18b takes over the magnetic influence and suppresses the magnetic dynamo of WASP-18.}
\begin{table*}[ht]
\caption{\label{hpress} Properties of stars with hot Jupiter s with $T_{eff}$ in the range $6200-6600$ K.
We list the effective temperatures, stellar masses and radii, planet-star separations, chromospheric
activity indicator  $\log R_{HK}^\prime$ (from \citealp{Knutson2010}), 
pressure scale heights ($H_P$), tidal heights ($H_t$) and 
ratio $H_t/H_P$ based on the formulae given by \citet{Cuntz2000}. 
Stellar data are taken from exoplanets.eu catalog and ordered by decreasing $H_t/H_P$ ratio. 
Data of WASP-19 are shown because of the planet-star separation very similar to WASP-18.
WASP-18 is the only star with pressure scale height and tidal height of the same order of magnitude. 
The second highest $H_t/H_P$ ratio ($\sim 20\%$) is seen in WASP-12, which has a low activity like WASP-18.}
\centering
\resizebox{0.68\textwidth}{!}{
\begin{tabular}{llllllllll}
  \hline\hline
 Star & T$_{eff}$ & R$_{star}$ & M$_{star}$ & M$_{planet}$ & Separation & $\log R_{HK}^\prime$ & $H_P$ & $H_t$ & $H_t/H_P$ \\ 
    & K & R$_\odot$ & M$_\odot$ & M$_{Jup}$ & AU & & km & km & \\
  \hline
WASP-18 & 6400 & 1.29 & 1.28 & 10.43 & 0.02047 & -5.43 & 419 & 498.3 & 1.189 \\
WASP-12 & 6300 & 1.599 & 1.35 & 1.404 & 0.02293 & -5.5 & 600.1 & 122.3 & 0.204 \\ 
 WASP-14 & 6475 & 1.306 & 1.211 & 7.341 & 0.036 & -4.923 & 458.7 & 44 & 0.096 \\ 
XO-3 & 6429 & 1.377 & 1.213 & 11.79 & 0.0454 & -4.595 & 505.5 & 39.4 & 0.078 \\ 
HAT-P-7 & 6350 & 1.84 & 1.47 & 1.8 & 0.0379 & -5.018 & 735.5 & 37.2 & 0.051 \\ 
HAT-P-2 & 6290 & 1.64 & 1.36 & 8.74 & 0.0674 & -4.78 & 625.6 & 14.6 & 0.023 \\ 
Kepler-5 & 6297 & 1.793 & 1.374 & 2.114 & 0.05064 & -5.037 & 740.9 & 14.1 & 0.019 \\ 
 HAT-P-14 & 6600 & 1.468 & 1.386 & 2.2 & 0.0594 & -4.855 & 516 & 3.4 & 0.007 \\ 
 HAT-P-6 & 6570 & 1.46 & 1.29 & 1.057 & 0.05235 & -4.799 & 545.9 & 2.6 & 0.005 \\ 
Kepler-8 & 6213 & 1.486 & 1.213 & 0.603 & 0.0483 & -5.05 & 568.8 & 2.3 & 0.004 \\ 
 WASP-17 & 6650 & 1.38 & 1.2 & 0.486 & 0.0515 & -5.331 & 530.7 & 1.1 & 0.002 \\ 
HAT-P-9 & 6350 & 1.32 & 1.28 & 0.67 & 0.053 & -5.092 & 434.7 & 1 & 0.002 \\ 
   \hline
 WASP-19 & 5500 & 1.004 & 0.904 & 1.114 & 0.01616 & -4.66 & 308.5 & 55.2 & 0.179 \\ 
\hline
\end{tabular}
}
\end{table*}

In summary, the factors that lead to SPI are not only a function of planet-star separation
and planet/star mass ratio, but rely also on the inner structure of the parent star, 
the efficiency of its dynamo, its age and the the strenght of a planetary magnetic field.

%
\begin{table*}
\caption{\label{opt-id} Optical and IR counterparts of X-ray sources. The first part of the Table
lists the six NED counterparts, the second part lists the two {\em 2MASS} counterparts, the third
part lists the only Simbad counterpart.}
\resizebox{\textwidth}{!}{
\begin{tabular}{r r r r l r r l l r }\hline \hline
  \multicolumn{1}{c}{\# } &
  \multicolumn{1}{c}{RA (deg)} &
  \multicolumn{1}{c}{Dec (deg)} &
  \multicolumn{1}{c}{No.} &
  \multicolumn{1}{c}{NED name} &
  \multicolumn{1}{c}{RA(deg)} &
  \multicolumn{1}{c}{DEC(deg)} &
  \multicolumn{1}{c}{Type} &
  \multicolumn{1}{c}{Magnitude } &
  \multicolumn{1}{c}{Separation} \\
 &\multicolumn{2}{c}{X-ray pos.} & & & & & & and Filter & arcsec \\ 
\hline
 20 & 24.29519 & -45.57693 & 62 & APMUKS(BJ) B013505.04-454952.2 & 24.29543 & -45.57682 & G & 18.99 & 0.7 \\
 22 & 24.3018 & -45.60976 & 30 & MRSS 244-010213 & 24.30153 & -45.60971 & G & 18.9r & 0.7 \\
129 & 24.35544 & -45.66886 & 6 & APMUKS(BJ) B013519.66-455523.1 & 24.3557 & -45.66886 & G & 20.29 & 0.7 \\
131 & 24.3612 & -45.67118 & 5 & APMUKS(BJ) B013520.98-455532.3 & 24.36116 & -45.67142 & G & 19.67 & 0.9 \\
184 & 24.49494 & -45.65133 & 49 & MRSS 244-008106 & 24.49512 & -45.65108 & G & 18.0r & 1.0 \\
186 & 24.50386 & -45.75107 & 70 & APMUKS(BJ) B013555.40-460017.6 & 24.5036 & -45.75096 & G & 19.55 & 0.8\\
\hline\end{tabular}
}
\resizebox{\textwidth}{!}{
\begin{tabular}{r r r r r l r r r r r r l r }
  \multicolumn{1}{c}{\# } &
  \multicolumn{1}{c}{RA} &
  \multicolumn{1}{c}{Dec} &
  \multicolumn{1}{c}{RAJ2000} &
  \multicolumn{1}{c}{DEJ2000} &
  \multicolumn{1}{c}{2MASS name} &
  \multicolumn{1}{c}{Jmag} &
  \multicolumn{1}{c}{e\_Jmag} &
  \multicolumn{1}{c}{Hmag} &
  \multicolumn{1}{c}{e\_Hmag} &
  \multicolumn{1}{c}{Kmag} &
  \multicolumn{1}{c}{e\_Kmag} &
  \multicolumn{1}{c}{Qflg} &
  \multicolumn{1}{c}{Sep.} \\
\hline
 20 &  24.29519 & -45.57693 & 24.295321 & -45.577065 & 01371087-4534374 & 16.6 & 0.14 & 15.9 & 0.13 & 15.2 & 0.16 & BBC & 0.6 \\
184 &  24.49494 & -45.65133 & 24.494989 & -45.65136 & 01375879-4539048 & 16.74 & 0.154 & 15.80 & 0.15 & 15.0 & 0.13 & BBB & 0.2 \\
\hline\end{tabular}
}
\resizebox{\textwidth}{!}{
\begin{tabular}{r  r r l l r r r r r r l r }
  \multicolumn{1}{c}{\# } &
  \multicolumn{1}{c}{RA} &
  \multicolumn{1}{c}{DEC} &
  \multicolumn{1}{c}{Simbad~ID} &
  \multicolumn{1}{c}{OTYPE} &
  \multicolumn{1}{c}{pmRA} &
  \multicolumn{1}{c}{pmDEC} &
  \multicolumn{1}{c}{Plx\_VALUE} &
  \multicolumn{1}{c}{Z\_VALUE} &
  \multicolumn{1}{c}{B} &
  \multicolumn{1}{c}{V} &
  \multicolumn{1}{c}{SP\_TYPE} &
  \multicolumn{1}{c}{Sep.} \\
\hline
211 & 24.743367 & -45.774851 & HD  10210 & Star & 48.35 & -1.45 & 6.33 & 4.8E-5 & 9.02 & 8.08 & G8III/IV & 1.6\\
\hline\end{tabular}
}
\end{table*}

\begin{table*}
\caption{\label{fit} Best fit models and parameters. We used absorbed powe law as model (Abs+Pow) 
for all but two cases, where we used absorbed bremsstrahlung (src. \# 64, Abs+Brems) and two thermal
modela (src. \# 211, Apec+Apec). Errors are  given at $1\sigma$ level.}
\resizebox{\textwidth}{!}{
\begin{tabular}{ccccccccccccc}\hline\hline
\#  & Model & $\chi^2$ & D.o.F. & N$_H$ & err$(N_H)$ & kT/$\alpha$ & err$(KT/\alpha)$ & Norm & err(Norm) &  & & flux \\
    &  name &          &        & \multicolumn{2}{c}{cm$^{-2}$} & \multicolumn{2}{c}{keV/--} & \multicolumn{2}{c}{cm$^{-5}$} & & & erg s$^{-1}$ cm$^{-2}$ \\\hline
  8 & Abs+Pow & 2.82    & 6   & 0.10  & 0.22   & 1.8  & 0.5     &  4.2e-06   & 2.0e-06 & & & 2.0e-14 \\
 20 & Abs+Pow & 4.13    & 4   & 0.35  & 0.50   & 0.51 & 0.5     &  1.8e-06   & 1.3e-06 & & & 4.0e-14 \\
22  & Abs+Pow & 4.13    &  4  & 0.35  & 0.5    & 0.5  & 0.5     &  1.8e-6    & 1.2e-6  & & & 5.4e-14 \\
 57 & Abs+Pow & 9.7     & 5   & 0.4   & 0.3    & 3.5  & 0.8     &  1.7e-05   & 1.0e-05 & & & 1.7e-14 \\
64  & Abs+Brems& 13.0   & 10  & 0.61  & 0.27   & 4.9  & 2.9     &  1.33e-05  & 4.1e-06 & & & 3.5e-14 \\
90  & Abs+Pow & 6.91    & 3   & 4.6   & 1.9    & 2.9  & 1.0     &  4.3e-05   & 5.8e-05 & & & 1.8e-14 \\
104 & Abs+Pow & 4.51    & 6   & 0.13  & 0.16   & 1.9  & 0.4     &  5.5e-06   & 2.1e-06 & & & 2.2e-14 \\
135 & Abs+Pow & 0.45    & 1   & 1.2   & 0.8    & 2.1  & 0.8     &  4.4e-06   & 4.8e-06 & & & 9.6e-15 \\
145 & Abs+Pow & 0.96    & 1   & 0.02  & 0.20   & 1.8  & 0.5     &  1.6e-06   & 0.9e-06 & & & 9.0e-15 \\
182 & Abs+Pow & 2.35    & 2   & 0.0   & 0.2    & 1.6  & 0.5     &  1.7e-06   & 0.9e-06 & & & 1.2e-14 \\
190 & Abs+Pow & 0.22    & 3   & 0.0   & 0.23   & 1.5  & 0.5     &  4.3e-06   & 2.3e-06 & & & 3.1e-14 \\\hline
\#  & Model & $\chi^2$ & D.o.F. & $kT_1$ & err$(KT_!)$ & $kT_2$ & err$(KT_2)$ & Norm1 & err(Norm1) & Norm2 & err(Norm2) &flux \\
211 & Apec+Apec& 41.1   & 36  & 0.37 & 0.06    & 0.75 & 0.1     &  4.1e-5    & 2.0e-5  & 2.7e-5 & 1.7e-5 & 9.1e-14 \\
\hline
\end{tabular} 
}
\end{table*}

\begin{figure}
\begin{center}
\includegraphics[width=0.9\columnwidth,angle=0]{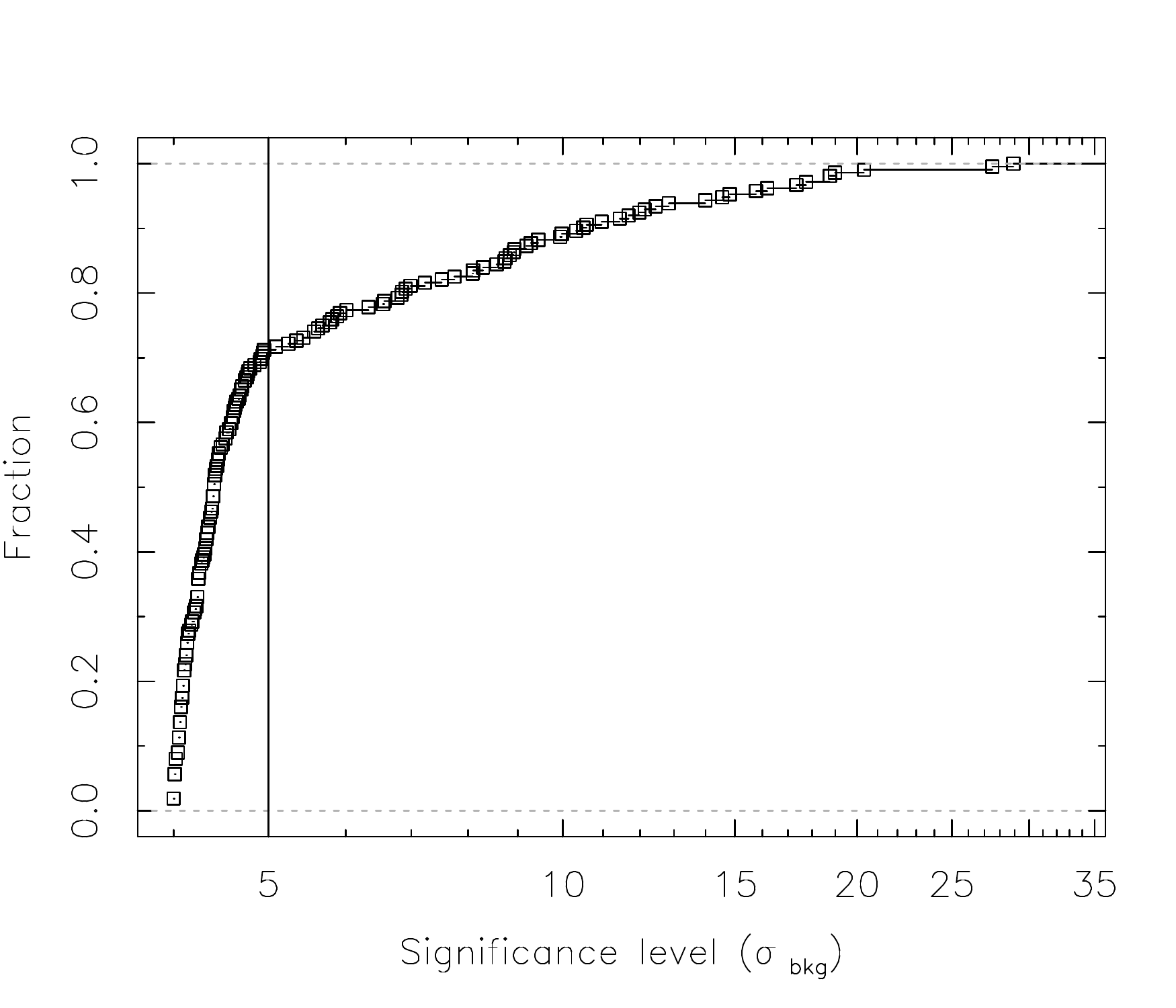}
\end{center}
\caption{\label{signif_dist} Cumulative distribution of significance in $\sigma_{bkg}$. Scale on x-axis is log.
A break at a significance level $\sigma=5$ is noticed.} 
\end{figure}

\begin{figure}
\begin{center}
\includegraphics[width=0.99\columnwidth,angle=0]{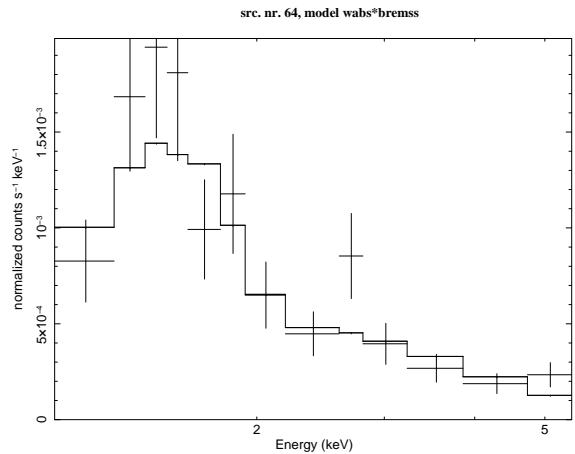}
\end{center}
\caption{\label{spec_64} X-ray spectrum and best fit model of src. 64. A line at $\sim2.5$ keV is
noticed.}
\end{figure}

\begin{figure}
\begin{center}
\includegraphics[width=0.99\columnwidth,angle=0]{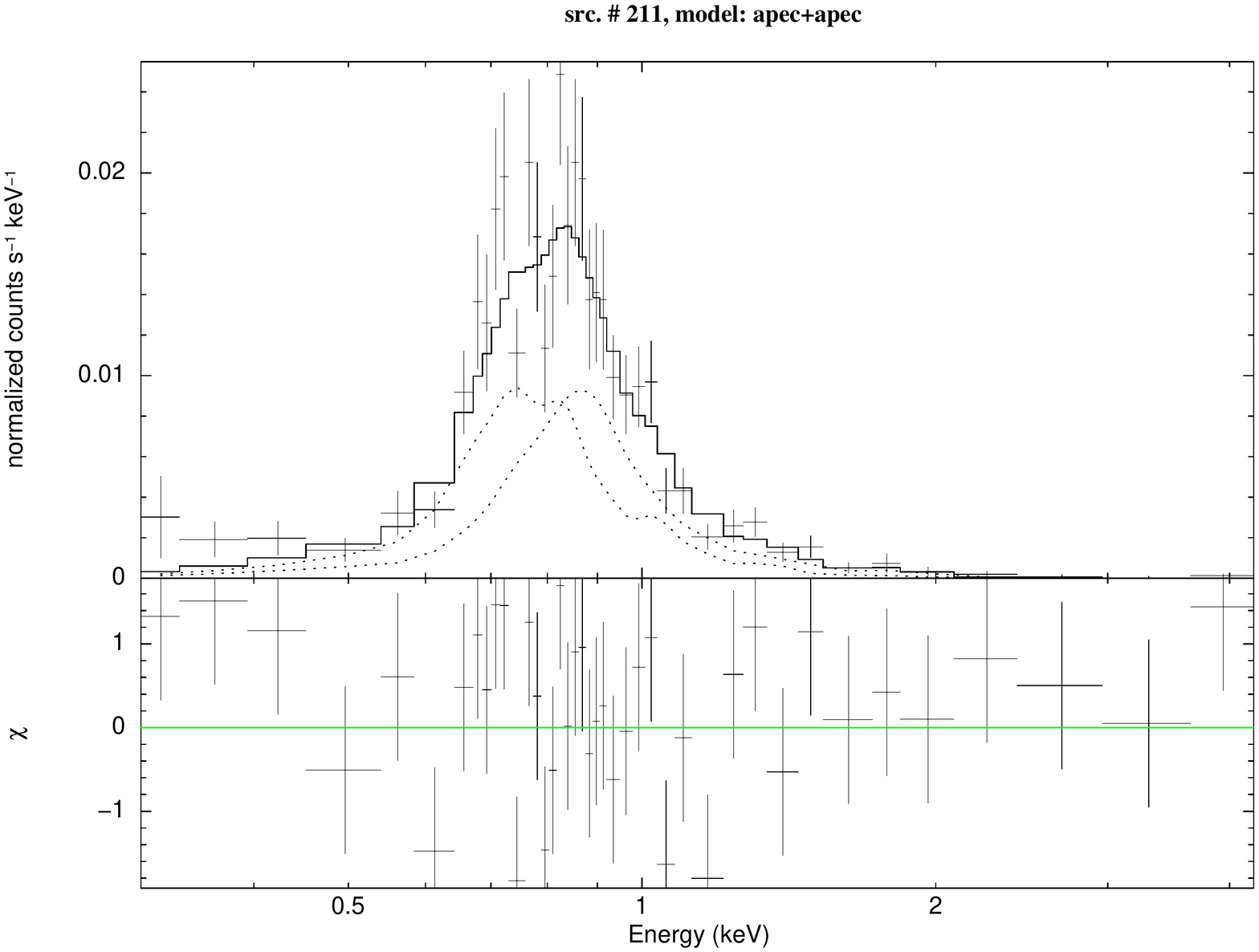}
\end{center}
\caption{\label{spec_211} X-ray spectrum and best fit model of the G8III/IV star HD~10210 (src. 211). 
A thermal model with two components ($kT_1 = 0.37$ keV, $KT_2 = 0,75$) is the best fit to the spectrum.}
\end{figure}

\section{Summary and conclusions}
Aimed to detect effects of star-planet interaction at high energies, 
we have analyzed a 87 ks deep \chandra\ observation pointed toward the star with hot Jupiter
WASP-18. We do not find X-ray emission from the star at a level above 
$L_X = 4.5 \times 10^{26}$ erg s$^{-1}$.
The star is at least 2.5 orders of magnitude less luminous in X-rays than analogs F late stars 
in Hyades, and  main sequence stars like $\tau$ Boo, the 4 Gyr old star HD~110450
and Procyon, which is at the very end of the main sequence or already post main sequence.
The absence of X-ray activity is in agreement with the low chromospheric activity reported by 
\citet{Knutson2010}, with the absence of reversal core emission in H$\alpha$ and Ca H\&K line. 
These facts strongly conflict with the estimate of the age from fitting to isochrones (600 Myr) 
given by \citet{Hellier2009} and \citet{Southworth2009}, and the strong Li absorption observed 
in optical spectra and would suggest an activity level more consistent with a solar age. 

A stellar age of a few Gyr puts strong constraints on the evolution model of the planet. 
In particular, it implies that the inward planet migration took place on a time scale of a few Gyr, 
not on time scale of hundreds of Myr as assessed by \citet{Brown2011}. 
Hence, the orbital evolution of WASP-18~b may have been slower than predicted by models. 

The absence of X-rays activity in the star indicates also a null efficiency of the 
magnetic dynamo. In these conditions, magnetic SPI is not at work, but
rather, a strong tidal influence from the massive hot Jupiter can have a major role in determining 
the outer stellar structure and activity of WASP-18. To reconcile the strong Li absorption 
with an absent activity, we suggest a scenario in which the tidal interaction of the massive planet 
has modified the inner stellar mixing, thus preventing or at least reducing the Li burning.
At the same time, the upper layers of the thin convective zone expected at this stellar mass are 
profoundly altered by tidal stresses.
In a sample of stars with hot Jupiters in the same range of effective temperature,
WASP-18 is the only object to show a tide height, induced by its planet, higher than the gas 
pressure scale height. The motions induced by tidal SPI could reduce the shear within
the convective zone, hampering the creation of a magnetic dynamo and thus the outer corona and the
production of X-rays. 
{ The same tides could be responsible for a reduction of the mixing efficiency in the inner
stellar layers resulting in a higher Li abundance than observed in stars of similar mass, 
like $\tau$~Boo.} 
This hypothesis requires detailed simulations in the framework of two bodies interaction
and modification of the stellar structure in binary systems.  
Our results can be a stimulus to understand any of these effects, and highlight the uniqueness of
WASP-18 among systems with hot Jupiters.

We find 212 X-ray sources in the ACIS image. We briefly discuss their characteristics in the
Appendix.

\begin{acknowledgements}
IP and SJW are grateful to dr. S. Saar for his comments on this paper.
IP acknowledges financial support of the European Union
under the project ``Astronomy Fellowships in Italy" (AstroFit).
S.J.W. was supported by NASA contract NAS8-03060.
VA is  grateful to dr. R. Trampedach for the discussion about the models of stellar convection.
VA acknowledges funding for the Stellar Astrophysics Centre provided by The Danish National 
Research Foundation; the research is supported by the ASTERISK project 
(ASTERoseismic Investigations with SONG and Kepler) funded by the European Research 
Council (Grant agreement no. 267864). 
\end{acknowledgements}


\begin{thebibliography}{62}
\expandafter\ifx\csname natexlab\endcsname\relax\def\natexlab#1{#1}\fi

\bibitem[{{Baliunas} {et~al.}(1995){Baliunas}, {Donahue}, {Soon}, {Horne},
  {Frazer}, {Woodard-Eklund}, {Bradford}, {Rao}, {Wilson}, {Zhang}, {Bennett},
  {Briggs}, {Carroll}, {Duncan}, {Figueroa}, {Lanning}, {Misch}, {Mueller},
  {Noyes}, {Poppe}, {Porter}, {Robinson}, {Russell}, {Shelton}, {Soyumer},
  {Vaughan}, \& {Whitney}}]{Baliunas1995}
{Baliunas}, S.~L., {Donahue}, R.~A., {Soon}, W.~H., {et~al.} 1995, \apj, 438,
  269

\bibitem[{{Bouvier}(2008)}]{Bouvier2008}
{Bouvier}, J. 2008, \aap, 489, L53

\bibitem[{Brown {et~al.}(2011)Brown, Cameron, Hall, Hebb, \&
  Smalley}]{Brown2011}
Brown, D. J.~A., Cameron, A.~C., Hall, C., Hebb, L., \& Smalley, B. 2011,
  Monthly Notices of the Royal Astronomical Society, 415, 605

\bibitem[{{Casagrande} {et~al.}(2011){Casagrande}, {Sch{\"o}nrich}, {Asplund},
  {Cassisi}, {Ram{\'{\i}}rez}, {Mel{\'e}ndez}, {Bensby}, \&
  {Feltzing}}]{Casagrande2011}
{Casagrande}, L., {Sch{\"o}nrich}, R., {Asplund}, M., {et~al.} 2011, \aap, 530,
  A138

\bibitem[{{Catala} {et~al.}(2007){Catala}, {Donati}, {Shkolnik}, {Bohlender},
  \& {Alecian}}]{Catala2007}
{Catala}, C., {Donati}, J.-F., {Shkolnik}, E., {Bohlender}, D., \& {Alecian},
  E. 2007, \mnras, 374, L42

\bibitem[{{Claret}(2004)}]{Claret2004}
{Claret}, A. 2004, \aap, 424, 919

\bibitem[{{Claret}(2005)}]{Claret2005}
{Claret}, A. 2005, \aap, 440, 647

\bibitem[{{Claret}(2006)}]{Claret2006}
{Claret}, A. 2006, \aap, 453, 769

\bibitem[{{Claret}(2007)}]{Claret2007}
{Claret}, A. 2007, \aap, 467, 1389

\bibitem[{{Cuntz} {et~al.}(2000){Cuntz}, {Saar}, \& {Musielak}}]{Cuntz2000}
{Cuntz}, M., {Saar}, S.~H., \& {Musielak}, Z.~E. 2000, \apjl, 533, L151

\bibitem[{{Damiani} {et~al.}(2003){Damiani}, {Flaccomio}, {Micela},
  {Sciortino}, {Harnden}, {Murray}, {Wolk}, \& {Jeffries}}]{Damiani2003}
{Damiani}, F., {Flaccomio}, E., {Micela}, G., {et~al.} 2003, \apj, 588, 1009

\bibitem[{{Damiani} {et~al.}(1997{\natexlab{a}}){Damiani}, {Maggio}, {Micela},
  \& {Sciortino}}]{Damiani97b}
{Damiani}, F., {Maggio}, A., {Micela}, G., \& {Sciortino}, S.
  1997{\natexlab{a}}, \apj, 483, 350

\bibitem[{{Damiani} {et~al.}(1997{\natexlab{b}}){Damiani}, {Maggio}, {Micela},
  \& {Sciortino}}]{Damiani97a}
{Damiani}, F., {Maggio}, A., {Micela}, G., \& {Sciortino}, S.
  1997{\natexlab{b}}, \apj, 483, 370

\bibitem[{{Delgado Mena} {et~al.}(2014){Delgado Mena}, {Israelian},
  {Gonz{\'a}lez Hern{\'a}ndez}, {Sousa}, {Mortier}, {Santos}, {Adibekyan},
  {Fernandes}, {Rebolo}, {Udry}, \& {Mayor}}]{DelgadoMena2014}
{Delgado Mena}, E., {Israelian}, G., {Gonz{\'a}lez Hern{\'a}ndez}, J.~I.,
  {et~al.} 2014, \aap, 562, A92

\bibitem[{{Deliyannis} {et~al.}(1994){Deliyannis}, {King}, {Boesgaard}, \&
  {Ryan}}]{Deliyannis1994}
{Deliyannis}, C.~P., {King}, J.~R., {Boesgaard}, A.~M., \& {Ryan}, S.~G. 1994,
  \apjl, 434, L71

\bibitem[{{Demarque} {et~al.}(2004){Demarque}, {Woo}, {Kim}, \&
  {Yi}}]{Demarque2004}
{Demarque}, P., {Woo}, J.-H., {Kim}, Y.-C., \& {Yi}, S.~K. 2004, \apjs, 155,
  667

\bibitem[{{Doyle} {et~al.}(2013){Doyle}, {Smalley}, {Maxted}, {Anderson},
  {Cameron}, {Gillon}, {Hellier}, {Pollacco}, {Queloz}, {Triaud}, \&
  {West}}]{Doyle2013}
{Doyle}, A.~P., {Smalley}, B., {Maxted}, P.~F.~L., {et~al.} 2013, \mnras, 428,
  3164

\bibitem[{{Eldridge} \& {Tout}(2004)}]{Eldridge2004}
{Eldridge}, J.~J. \& {Tout}, C.~A. 2004, \mnras, 353, 87

\bibitem[{{Erkaev} {et~al.}(2007){Erkaev}, {Kulikov}, {Lammer}, {Selsis},
  {Langmayr}, {Jaritz}, \& {Biernat}}]{Erkaev2007}
{Erkaev}, N.~V., {Kulikov}, Y.~N., {Lammer}, H., {et~al.} 2007, \aap, 472, 329

\bibitem[{{Fares} {et~al.}(2010){Fares}, {Donati}, {Moutou}, {Jardine},
  {Grie{\ss}meier}, {Zarka}, {Shkolnik}, {Bohlender}, {Catala}, \& {Collier
  Cameron}}]{Fares2010}
{Fares}, R., {Donati}, J.-F., {Moutou}, C., {et~al.} 2010, \mnras, 406, 409

\bibitem[{{Giardino} {et~al.}(2007){Giardino}, {Favata}, {Pillitteri},
  {Flaccomio}, {Micela}, \& {Sciortino}}]{Giardino07}
{Giardino}, G., {Favata}, F., {Pillitteri}, I., {et~al.} 2007, \aap, 475, 891

\bibitem[{{Gonzalez}(2008)}]{Gonzalez2008}
{Gonzalez}, G. 2008, \mnras, 386, 928

\bibitem[{{Hellier} {et~al.}(2009){Hellier}, {Anderson}, {Collier Cameron},
  {Gillon}, {Hebb}, {Maxted}, {Queloz}, {Smalley}, {Triaud}, {West}, {Wilson},
  {Bentley}, {Enoch}, {Horne}, {Irwin}, {Lister}, {Mayor}, {Parley}, {Pepe},
  {Pollacco}, {Segransan}, {Udry}, \& {Wheatley}}]{Hellier2009}
{Hellier}, C., {Anderson}, D.~R., {Collier Cameron}, A., {et~al.} 2009, \nat,
  460, 1098

\bibitem[{{Holmberg} {et~al.}(2009){Holmberg}, {Nordstr{\"o}m}, \&
  {Andersen}}]{Holmberg2009}
{Holmberg}, J., {Nordstr{\"o}m}, B., \& {Andersen}, J. 2009, \aap, 501, 941

\bibitem[{{Houdek} {et~al.}(1999){Houdek}, {Balmforth},
  {Christensen-Dalsgaard}, \& {Gough}}]{Houdek1999}
{Houdek}, G., {Balmforth}, N.~J., {Christensen-Dalsgaard}, J., \& {Gough},
  D.~O. 1999, \aap, 351, 582

\bibitem[{{Ip} {et~al.}(2004){Ip}, {Kopp}, \& {Hu}}]{Ip2004}
{Ip}, W.-H., {Kopp}, A., \& {Hu}, J.-H. 2004, \apjl, 602, L53

\bibitem[{{Israelian} {et~al.}(2004){Israelian}, {Santos}, {Mayor}, \&
  {Rebolo}}]{Israelian2004}
{Israelian}, G., {Santos}, N.~C., {Mayor}, M., \& {Rebolo}, R. 2004, \aap, 414,
  601

\bibitem[{{Kashyap} {et~al.}(2008){Kashyap}, {Drake}, \& {Saar}}]{Kashyap08}
{Kashyap}, V.~L., {Drake}, J.~J., \& {Saar}, S.~H. 2008, \apj, 687, 1339

\bibitem[{{Knutson} {et~al.}(2010){Knutson}, {Howard}, \&
  {Isaacson}}]{Knutson2010}
{Knutson}, H.~A., {Howard}, A.~W., \& {Isaacson}, H. 2010, \apj, 720, 1569

\bibitem[{{Krej{\v c}ov{\'a}} \& {Budaj}(2012)}]{Krejcova2012}
{Krej{\v c}ov{\'a}}, T. \& {Budaj}, J. 2012, \aap, 540, A82

\bibitem[{{Melo} {et~al.}(2006){Melo}, {Santos}, {Pont}, {Guillot},
  {Israelian}, {Mayor}, {Queloz}, \& {Udry}}]{Melo2006}
{Melo}, C., {Santos}, N.~C., {Pont}, F., {et~al.} 2006, \aap, 460, 251

\bibitem[{{Miller} {et~al.}(2012){Miller}, {Gallo}, {Wright}, \&
  {Dupree}}]{Miller2012}
{Miller}, B.~P., {Gallo}, E., {Wright}, J.~T., \& {Dupree}, A.~K. 2012, \apj,
  754, 137

\bibitem[{{Pace} {et~al.}(2012){Pace}, {Castro}, {Mel{\'e}ndez}, {Th{\'e}ado},
  \& {do Nascimento}}]{Pace2012}
{Pace}, G., {Castro}, M., {Mel{\'e}ndez}, J., {Th{\'e}ado}, S., \& {do
  Nascimento}, Jr., J.-D. 2012, \aap, 541, A150

\bibitem[{{Pallavicini} {et~al.}(1981){Pallavicini}, {Golub}, {Rosner},
  {Vaiana}, {Ayres}, \& {Linsky}}]{Pallavicini81}
{Pallavicini}, R., {Golub}, L., {Rosner}, R., {et~al.} 1981, \apj, 248, 279

\bibitem[{{Penz} \& {Micela}(2008)}]{Penz08}
{Penz}, T. \& {Micela}, G. 2008, \aap, 479, 579

\bibitem[{{Pillitteri} {et~al.}(2011){Pillitteri}, {G{\"u}nther}, {Wolk},
  {Kashyap}, \& {Cohen}}]{Pillitteri2011}
{Pillitteri}, I., {G{\"u}nther}, H.~M., {Wolk}, S.~J., {Kashyap}, V.~L., \&
  {Cohen}, O. 2011, \apjl, 741, L18

\bibitem[{{Pillitteri} {et~al.}(2010){Pillitteri}, {Wolk}, {Cohen}, {Kashyap},
  {Knutson}, {Lisse}, \& {Henry}}]{Pillitteri2010}
{Pillitteri}, I., {Wolk}, S.~J., {Cohen}, O., {et~al.} 2010, \apj, 722, 1216

\bibitem[{{Pillitteri} {et~al.}(2014){Pillitteri}, {Wolk}, {Lopez-Santiago},
  {Guenther}, {Sciortino}, {Cohen}, {Kashyap}, \& {Drake}}]{Pillitteri2014}
{Pillitteri}, I., {Wolk}, S.~J., {Lopez-Santiago}, J., {et~al.} 2014, ArXiv
  e-prints

\bibitem[{{Pizzolato} {et~al.}(2003){Pizzolato}, {Maggio}, {Micela},
  {Sciortino}, \& {Ventura}}]{Pizzolato2003}
{Pizzolato}, N., {Maggio}, A., {Micela}, G., {Sciortino}, S., \& {Ventura}, P.
  2003, \aap, 397, 147

\bibitem[{{Pols} {et~al.}(1998){Pols}, {Schr{\"o}der}, {Hurley}, {Tout}, \&
  {Eggleton}}]{Pols1998}
{Pols}, O.~R., {Schr{\"o}der}, K.-P., {Hurley}, J.~R., {Tout}, C.~A., \&
  {Eggleton}, P.~P. 1998, \mnras, 298, 525

\bibitem[{{Poppenhaeger} {et~al.}(2010){Poppenhaeger}, {Robrade}, \&
  {Schmitt}}]{Poppenhaeger2010}
{Poppenhaeger}, K., {Robrade}, J., \& {Schmitt}, J.~H.~M.~M. 2010, \aap, 515,
  A98+

\bibitem[{{Poppenhaeger} \& {Schmitt}(2011)}]{Poppenhaeger2011}
{Poppenhaeger}, K. \& {Schmitt}, J.~H.~M.~M. 2011, \apj, 735, 59

\bibitem[{{Poppenhaeger} {et~al.}(2013){Poppenhaeger}, {Schmitt}, \&
  {Wolk}}]{Poppenhaeger2013}
{Poppenhaeger}, K., {Schmitt}, J.~H.~M.~M., \& {Wolk}, S.~J. 2013, \apj, 773,
  62

\bibitem[{{Poppenhaeger} \& {Wolk}(2014)}]{Poppenhaeger2014}
{Poppenhaeger}, K. \& {Wolk}, S.~J. 2014, ArXiv e-prints

\bibitem[{{Randich} \& {Schmitt}(1995)}]{Randich1995}
{Randich}, S. \& {Schmitt}, J.~H.~M.~M. 1995, \aap, 298, 115

\bibitem[{{Randich} {et~al.}(1998){Randich}, {Singh}, {Simon}, {Drake}, \&
  {Schmitt}}]{Randich1998}
{Randich}, S., {Singh}, K.~P., {Simon}, T., {Drake}, S.~A., \& {Schmitt},
  J.~H.~M.~M. 1998, \aap, 337, 372

\bibitem[{{Saar} {et~al.}(2004){Saar}, {Cuntz}, \& {Shkolnik}}]{Saar2004}
{Saar}, S.~H., {Cuntz}, M., \& {Shkolnik}, E. 2004, in IAU Symposium, Vol. 219,
  Stars as Suns : Activity, Evolution and Planets, ed. {A.~K.~Dupree \&
  A.~O.~Benz}, 355--+

\bibitem[{{Sanz-Forcada} {et~al.}(2011){Sanz-Forcada}, {Micela}, {Ribas},
  {Pollock}, {Eiroa}, {Velasco}, {Solano}, \&
  {Garc{\'{\i}}a-{\'A}lvarez}}]{Sanz-Forcada2011}
{Sanz-Forcada}, J., {Micela}, G., {Ribas}, I., {et~al.} 2011, \aap, 532, A6

\bibitem[{{Schr{\"o}ter} {et~al.}(2011){Schr{\"o}ter}, {Czesla}, {Wolter},
  {M{\"u}ller}, {Huber}, \& {Schmitt}}]{Schroeter2011}
{Schr{\"o}ter}, S., {Czesla}, S., {Wolter}, U., {et~al.} 2011, \aap, 532, A3+

\bibitem[{{Shkolnik} {et~al.}(2003){Shkolnik}, {Walker}, \&
  {Bohlender}}]{Shkolnik03}
{Shkolnik}, E., {Walker}, G.~A.~H., \& {Bohlender}, D.~A. 2003, \apj, 597, 1092

\bibitem[{{Skumanich}(1972)}]{Skumanich1972}
{Skumanich}, A. 1972, \apj, 171, 565

\bibitem[{{Soderblom}(2010)}]{Soderblom2010}
{Soderblom}, D.~R. 2010, \araa, 48, 581

\bibitem[{{Southworth} {et~al.}(2009){Southworth}, {Hinse}, {Dominik},
  {Glitrup}, {J{\o}rgensen}, {Liebig}, {Mathiasen}, {Anderson}, {Bozza},
  {Browne}, {Burgdorf}, {Calchi Novati}, {Dreizler}, {Finet}, {Harps{\o}e},
  {Hessman}, {Hundertmark}, {Maier}, {Mancini}, {Maxted}, {Rahvar}, {Ricci},
  {Scarpetta}, {Skottfelt}, {Snodgrass}, {Surdej}, \&
  {Zimmer}}]{Southworth2009}
{Southworth}, J., {Hinse}, T.~C., {Dominik}, M., {et~al.} 2009, \apj, 707, 167

\bibitem[{{Stern} {et~al.}(1995){Stern}, {Schmitt}, \& {Kahabka}}]{Stern1995}
{Stern}, R.~A., {Schmitt}, J.~H.~M.~M., \& {Kahabka}, P.~T. 1995, \apj, 448,
  683

\bibitem[{{Takeda} {et~al.}(2012){Takeda}, {Honda}, {Ohnishi}, {Ohkubo},
  {Hirata}, \& {Sadakane}}]{Takeda2012}
{Takeda}, Y., {Honda}, S., {Ohnishi}, T., {et~al.} 2012, ArXiv e-prints

\bibitem[{{Thorburn} {et~al.}(1993){Thorburn}, {Hobbs}, {Deliyannis}, \&
  {Pinsonneault}}]{Thorburn1993}
{Thorburn}, J.~A., {Hobbs}, L.~M., {Deliyannis}, C.~P., \& {Pinsonneault},
  M.~H. 1993, \apj, 415, 150

\bibitem[{{Torres} {et~al.}(2008){Torres}, {Winn}, \& {Holman}}]{Torres2008}
{Torres}, G., {Winn}, J.~N., \& {Holman}, M.~J. 2008, \apj, 677, 1324

\bibitem[{{Trampedach} {et~al.}(2013){Trampedach}, {Asplund}, {Collet},
  {Nordlund}, \& {Stein}}]{Trampedach2013}
{Trampedach}, R., {Asplund}, M., {Collet}, R., {Nordlund}, {\AA}., \& {Stein},
  R.~F. 2013, \apj, 769, 18

\bibitem[{{Vaughan} \& {Preston}(1980)}]{Vaughan1980}
{Vaughan}, A.~H. \& {Preston}, G.~W. 1980, \pasp, 92, 385

\bibitem[{{Walker} {et~al.}(2008){Walker}, {Croll}, {Matthews}, {Kuschnig},
  {Huber}, {Weiss}, {Shkolnik}, {Rucinski}, {Guenther}, {Moffat}, \&
  {Sasselov}}]{Walker2008}
{Walker}, G.~A.~H., {Croll}, B., {Matthews}, J.~M., {et~al.} 2008, \aap, 482,
  691

\bibitem[{{Wilson}(1966)}]{Wilson1966}
{Wilson}, O.~C. 1966, \apj, 144, 695

\bibitem[{{Yi} {et~al.}(2001){Yi}, {Demarque}, {Kim}, {Lee}, {Ree}, {Lejeune},
  \& {Barnes}}]{Yi2001}
{Yi}, S., {Demarque}, P., {Kim}, Y.-C., {et~al.} 2001, \apjs, 136, 417

\end{thebibliography}

%

\appendix

\section{Nature of the X-ray sources}
In the ACIS image we have detected 212 X-ray sources with significance $>4\sigma$. 
We have cross matched their spatial positions with Simbad, NED, and 2MASS
catalogs within a positional separation of 2\arcsec, obtaining seven matches that are listed 
in Table \ref{opt-id}. All NED matches are galaxies, and the X-ray emission could be associated to 
the AGNs at their center. One match is the giant/sub-giant G8III/IV star (HD 10210, src \# 211) with V = 8.08. 
The two matches in 2MASS are again two galaxies in NED catalog.

We are left with 205 X-ray sources with unknown match in the above catalogs. 
Most of these sources are faint as shown by the cumulative distribution of the significance values 
(Fig. \ref{signif_dist}). The cumulative distribution has a change of slope at about $\sigma=5$, 
marking the brighter sample from the rest of the sources. The number of unidentified sources 
with $\sigma>5$ is 55, or the 27\% of the total sample.
A number of them could be distant AGNs.

For the brightest sources we did a model best fit to the spectra
with one absorbed powerlaw, suited in the case of AGNs and well describing the featureless spectra
that we observe in these sources. In the bright source \# 64 we find a good best fit with an absorbed 
bremsstrahlung, but the presence of a line at $\sim2.5$ keV is also noticed.

The G8III/IV type star HD~10210 is also detected as the brightest in the sample (source \# 211), 
The best fit of the spectrum of HD 10210 has two temperatures at $kT = 0.37$ and $kT=0.75$, with the cool
component weighting twice than the hot component in the spectrum. Overall, the spectrum is similar to
that of a mid active main sequence star.
Detecting X-ray emission in a evolved star off of the main sequence is worth to be noticed, 
its X-ray luminosity is $L_X \sim 1.1\times 10^{29}$ erg/s.

\end{document}